# Comparison between Artificial Neural Network and Adaptive Neuro-Fuzzy Inference System For The Baryon-to-Meson Ratios in Proton-Proton Collisions

D. M. Habashy and H. I. Lebda

Physics Department, Faculty of Education, Ain Shams University, Roxy, Cairo 11757, Egypt

**Abstract**

This article presents two systems that can simulate and predict Particles ratios ($\Lambda_C^+/D^0$, $\Xi_C^0/D^0$, $\Xi/\phi$ and $\Omega/\phi$ ) created in high energy proton-proton (pp) collisions as a function of transverse momentum ($P_T$) and the center-of-mass energy ($\sqrt{s}$). An adaptive neurofuzzy inference system (ANFIS) and an artificial neural network (ANN) system are the systems in discussion. The ANFIS and ANN simulation results for training particles ratios as evaluated with training data points revealed an excellent match to the experimental data. The ANFIS and ANN's prediction abilities were also tested using data points that were not included in training and they performed well. The results clearly show that these methods are capable of extracting collision information and that they are helpful. Also, ANFIS and ANN results were compared with additional theoretical results (PYTHIA (CR Mode), HERWIG7, PYTHIA, PYTHIA8 (Monash), and EPOS-LHC).  It is found that the ANFIS shows better performance and is trained more quickly than the ANN system and other theoretical models.

## Introduction

The Quark-Gluon-Plasma (QGP), a deconfined state of matter created in heavy-ion collisions at extreme energy densities, can be studied through charm creation [1-5]. Hard Parton Scattering processes produce heavy quarks in the early stages of a collision, and as they travel through the QCD medium, they interact with its constituents and so experience the medium's evolution. In heavy-ion collisions, the charming baryon-to-meson ratio is particularly sensitive to the QGP's charm hadronisation mechanisms. Indeed, coalescence or recombination is projected to a significant fraction of low and intermediate velocity charm and beauty quarks with the other light quarks in the medium. As a result, the $\Lambda_C^+/D^0$ ratio improves with regard to proton-proton collisions. Furthermore, the presence of light di-quark bound states in the QGP [1] is likely to result in a further increase. The interpretation of heavy-ion collision results necessitates extensive study in smaller systems as well: At the LHC energies, pp collisions provide the essential reference for measurements and allow the testing of PQCD predictions and hadronisation models. Cold-nuclear matter phenomena that are connected to the existence of nuclei in the colliding system and that could resemble final-state medium-related effects are routinely investigated using p-Pb collisions.

Artificial Intelligence approaches such as Artificial Neural Networks (ANN) [6], Fuzzy Logic (FL) [7], and Adaptive Neuro-Fuzzy Inference System (ANFIS) [7-9] have been widely employed for the prediction and modeling of complicated physical systems as efficient alternative tools. One of the specialised uses of these methods in physics is modeling high and ultra-energy physics (HEP and UHE) collisions [10-13]. Developing a mathematical model for a system can be difficult and time-consuming, as it frequently necessitates making assumptions and neglecting some system factors. ANFIS is a type of Artificial Intelligence (AI) that helps with prediction, modeling, and inference. ANFIS is a Fuzzy Inference System (FIS) that works within the context of adaptive networks. It combines the ideas of Neural Networks (NN) and Fuzzy Logic (FL) into a single framework that can learn to estimate non-linear functions and operates as a universal estimator. It was created in the early 1990s and is known as Takagi–Sugeno FIS. This model's learning networks are built on mathematical computations that can solve difficult problems.

The goal of this study is to compute the Baryon-to-Meson ratios of the particles ($\Lambda_C^+/D^0$, $\Xi_C^0/D^0$, $\Xi/\phi$, and $\Omega/\phi$ ) at different high energies formed from PP collisions using ANN and ANFIS simulation model. We compare our results to the experimental data and other theoretical models. The following is how this paper is structured. In Sec. 2, we briefly introduce the ANN model and the ANFIS model in Sec.3. The results and analysis are shown in Sec. 4. The conclusion is drawn in Sec. 5.

## 2 Artificial Neural Network (ANN)

The back propagation (BP) learning method, which is now the most common in engineering applications, was included in this study among the various kinds of ANN approaches that exist. An artificial Neural Network (ANN) is considered a group of models roused by biological neural networks. They consist of several simple processors (neurons) that operate in parallel with no central control.

Neurons are organized into specific structures, usually layered, and connected by weighting elements called synapses. An input layer, at least one hidden layer, and the output layer are always present. The nodes number in the input layer is proportional to the number of parameters present in the dataset, and the nodes number in the output layer depends on the solution domain of the problem. The weighted inputs are summed up at each node and then sent to that node to produce an output. The activation functions passed the output of the node from the input layer to the output layer. Neural networks determine the correlation between two variables by refreshing weights and biases using Levenberg–Marquardt optimization. Modeling using artificial neural networks varies by extracting the data needed to train the network, the hidden layers number, the number of neurons in each hidden layer, and the type of activation function to find the optimal structure. It involves three steps: creating a nice structure, maintaining the structure, and finally testing the neural network to get the best one [14–15]. The Levenberg–Marquardt optimization is used to adjust weight and bias variables in the back propagation network training function. The Levenberg–Marquardt algorithm is very well suited to neural network training, where the performance index is the mean squared error [14]. Mean squared error (MSE) and correlation coefficient (R) that determine network performance is formulated as follows:

$$MSE = \frac{1}{N} \sum_{i=1}^{N}(y_{iact} - y_{ipred})^2 \qquad (1)$$

$$R = \sqrt{1 - \frac{\sum_{i=1}^{N}(y_{iact} - y_{ipred})^2}{\sum_{i=1}^{N}(y_{iact})^2 - \frac{\sum_{i=1}^{N}(y_{ipred})^2}{N}}} \qquad (2)$$

Where, $y_{iact}$ is the actual value of the ith pattern, $y_{ipred}$ is the predicted value of the ith pattern and N is the number of data points.

## 3. Adaptive Neuro-Fuzzy Inference (ANFIS)

### 3.1 Fuzzy logic (FL)

Fuzzy logic (FL) contains several logical values which are the true value of the variable or the problem in the range of 0 and 1. Lotfi Zadeh, the designer of fuzzy logic, observed that, unlike computers, fuzzy logic can be executed in systems of different sizes and capabilities, going from small micro-controllers to large, networked, workstation-based control systems. It can likewise be executed in both software and hardware [16].

## 3.2 The architecture of the Fuzzy Logic System

The architecture consists of four different parts listed below, shown in Fig. 1.

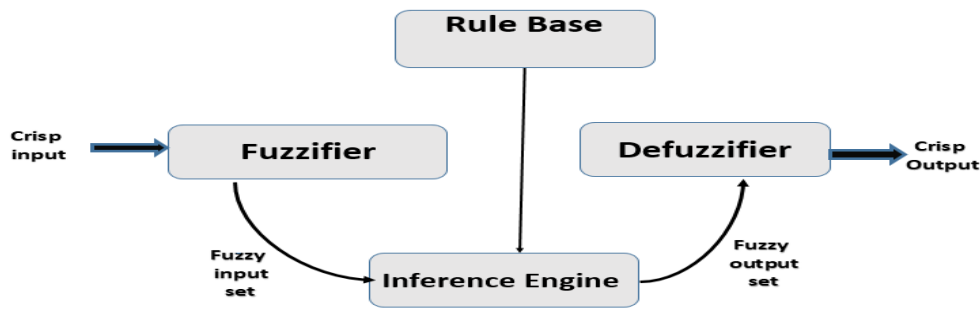

Fig.1 The fuzzy Logic system components

**The rule base** is where a set of rules is stored, and the If-Then conditions provided by the experts are utilized for controlling the decision-making systems. There are so many advancements in the Fuzzy theory as of late, which offers effective methods for planning and tuning fuzzy controllers. These advancements will reduce the number of fuzzy sets of rules.

**Fuzzification** is a part of the conversion of the system inputs, converting crisp quantities into fuzzy quantities.

**Inference Engine** is the most important aspect of any Fuzzy Logic System (FLS), as it is where all of the input is processed. The degree of matching between the rules and the current fuzzy input can be discovered by the user. Following the degree of matching, the system selects which rules to add based on the input field. After all of the rules have been executed, they are merged to form a control action.

**Defuzzification** is a part that takes the fuzzy set inputs generated by the Inference engine and transfers the fuzzy values into crisp values.

The fuzzy sets were presented by Zadeh [7] in 1965. The idea of fuzzy sets can be considered a generalization of ordinary (crisp) sets. The theory of fuzzy sets and the foundations of fuzzy logic were created by Zadeh dependent on the traditional classical logic and set theory.

Fuzzy sets form the basis of fuzzy logic and are completely determined by their membership function in the applications.

## 3.3 Membership Function

The membership function (MF) is a function that describes a fuzzy set's graph and allows users to quantify linguistic notions by defining how each point in the input space is converted into a membership value in the range of 0 to 1 [9,17]. Some membership functions are shown in Fig. 2.

Trimf. and trapmf., which are formed with straight lines, are the most basic membership functions. The trapezoidal membership function, trapmf, is a truncated triangular curve with a flat top. The Gaussian distribution curve is the basis for the gaussmf and gauss2mf functions. The terms gbellmf (generalised bell membership function), sigmf (sigmoidal membership function), and pimf (polynomial-based curves) are used interchangeably.

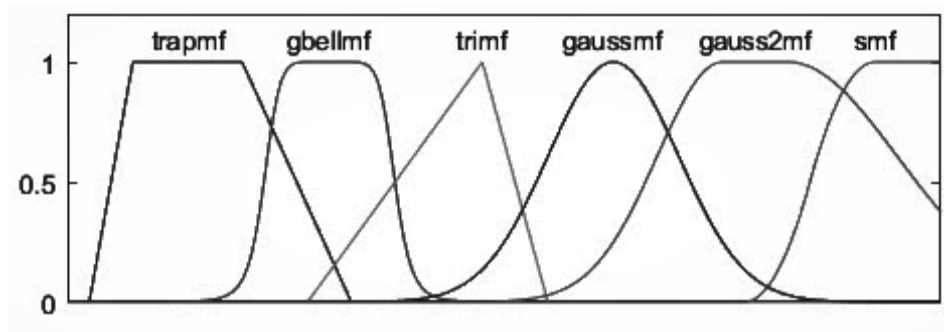

Fig. 2 Membership functions.

## 3.4 ANFIS Networks

One of the soft computing techniques that play a significant role in modelling accurate input-output matrix relationships is the adaptive neuro-fuzzy inference system (ANFIS) [9,18,21]. Neuro-Fuzzy Systems (NFS) are created by combining Neural Network (NN) and Fuzzy Logic (FL). The Adaptive Neuro-Fuzzy Inference System (ANFIS) was originally introduced by Jang in 1993 [22]. It is gotten from neuro-fuzzy systems, one of several types of fuzzy systems that are trained with a learning algorithm. It is called adaptive because it is trained to relearn its parameters (fuzzy sets and fuzzy rules) by processing data samples, in order to better represent input data. This kind of system can be utilized in fields such as pattern recognition, image processing, and signal processing.

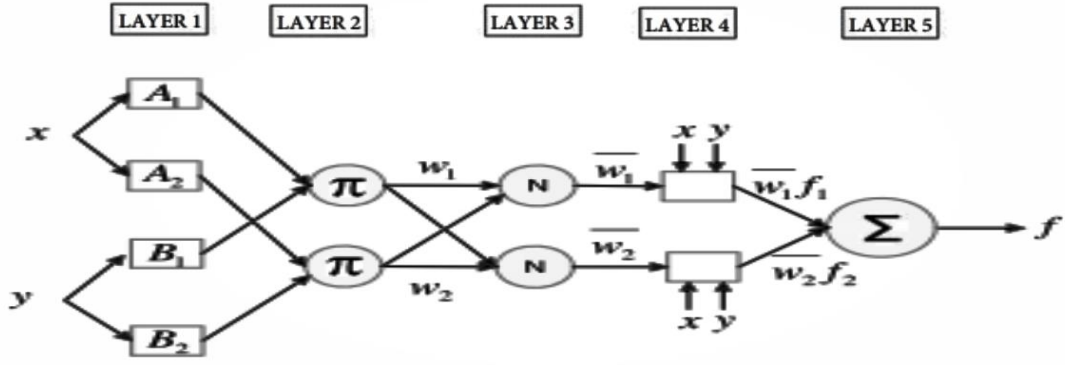

Fig.3 ANFIS architecture.

Fig. 3 depicts the architecture's five layers. The input values are received by the first layer, which determines the membership functions that apply to them. The "fuzzification layer" is what it's termed. This layer's outputs are the inputs' fuzzy membership grades, which are determined using the following equations:

$$O_{1,i} = \mu_{A_i}(x), \quad i = 1,2 \tag{3}$$

$$O_{1,i} = \mu_{B_i}(y), \quad i = 3,4 \tag{4}$$

Where $x$ and $y$ are the inputs to node i, and $A_i$ and $B_i$ are the linguistic labels associated with this node function. $\mu_{A_i}(x)$ and $\mu_{B_i}(y)$ can assume any fuzzy membership function.

The second layer is in charge of determining the rules' firing strengths. Because of its function, the second layer is known as the rule layer. Fuzzy operators are utilised in this layer. The AND operator is used to fuzzify the inputs. Layer 2's output can be stated as:

$$O_{2,i} = \omega_i = \mu_{A_i}(x) * \mu_{B_i}(y), \quad i = 1,2 \tag{5}$$

The third layer's job is to divide each value of the overall firing strength to normalise the computed firing strengths. The normalising layer is the name for this layer. Layer 3's output can be stated as:

$$O_{3,i} = \bar{\omega}_i = \frac{\omega_i}{\omega_1 + \omega_2}, \quad i = 1,2 \tag{6}$$

The normalised input values and the consequence parameter set are passed to the fourth layer.

$$O_{4,i} = \bar{\omega}_i f_i = \bar{\omega}_i (p_i x + q_i y + r_i), \quad i = 1,2 \tag{7}$$

Where $\bar{\omega}_i$ is the output of the third layer, and $p_i, q_i$ and $r_i$ are the consequent parameters.

The defuzzification values are returned by this layer, which is then passed on to the final layer, the summation layer, to generate the final output. The real output of ANFIS is obtained by adding the outputs obtained for each rule in the defuzzification layer.

The model's overall output is given by

$$O_{5,i} = \sum_i \overline{\omega}_i f_i = \frac{\sum_i \omega_i f_i}{\sum_i \omega_i}, \quad i = 1,2 \tag{8}$$

Creating, training, and testing the ANFIS model is done in the MATLAB software package. The MATLAB software package provides a powerful simulation and test platform used for this work, allowing easy manipulation of the model's variables and parameters. Therefore, you get a large graphic display of parameters and performance. Data for training or testing should be loaded into the anfis editor, create an initial FIS model, select the number of training epochs and training error tolerance, and select the FIS model parameter optimization method: backpropagation or a hybrid method. To train your FIS model, click the "Train Now" button. You can compare the FIS model output to the training, checking, or testing data output by clicking the Test Now button. To attain the best performance, we can change settings for items such as data set samples, epoch numbers, and membership function types. According to RMSE, the best training/ checking data set was selected. By comparing the difference between predicted and measured values, we calculated the prediction accuracy.

We have two inputs and one output. The input data is converted to degrees of memberships and membership values in a process called fuzzification. Each of the two inputs was divided into three triangular membership functions. Variables in both the input and output were imported into the ANFIS environment using the workspace button after clicking on load data. The fuzzification process was performed in the MATLAB FIS editor and the outcome is given in Fig. 4.

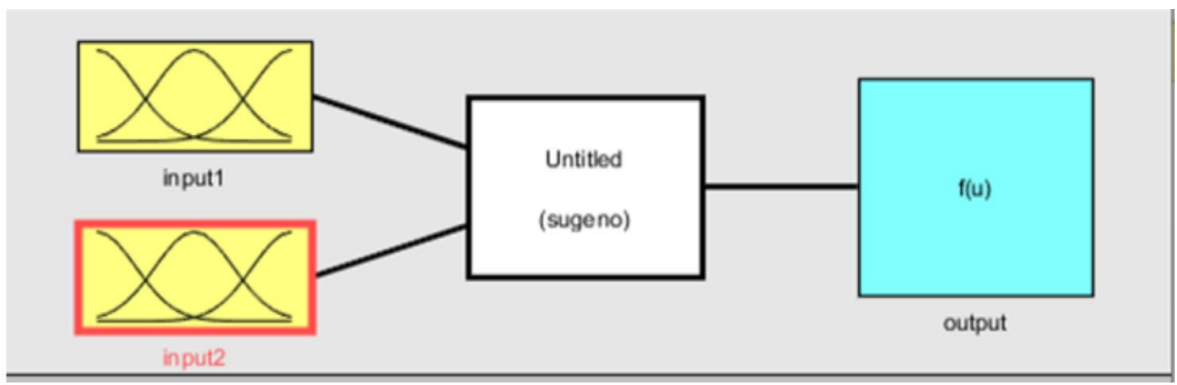

Fig.4: MATLAB Simulation ANFIS Model

## 3.5 ANN and ANFIS modeling for the Baryon-to-Meson ratios.

Data of p - p interactions at four particles ratios $\Lambda_C^+/D^0$, $\Xi_C^0/D^0$, $\Xi/\phi$ and $\Omega/\phi$ were considered as input to the ANN and ANFIS models. Accordingly, four different ANN and ANFIS models were proposed and their performances were compared to determine the best model. Four networks were used to simulate and predict particles ratios ($\Lambda_C^+/D^0$, $\Xi_C^0/D^0$, $\Xi/\phi$, $\Omega/\phi$ ) for p-p collisions as a function of $P_T$ at a wide range of C. M. energy $\sqrt{s}$ = 5.02 TeV, 7 TeV, and 13 TeV and mid-rapidity (|y| < 0.5). ANN and ANFIS models were trained with $P_T$ and $\sqrt{s}$ as inputs and different particles ratios as outputs for the same range of energy. Fig.5. show a simplification of the proposed ANN and ANFIS models.

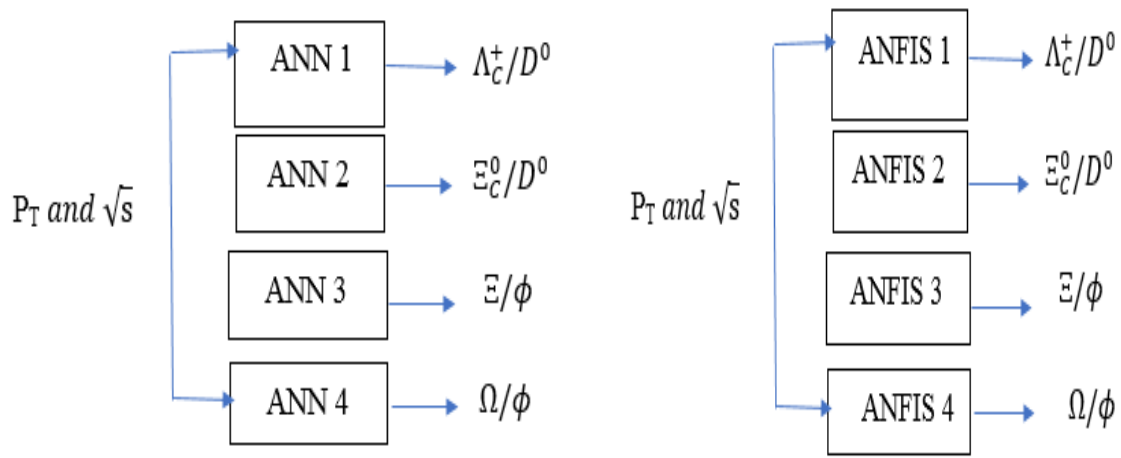

Fig.5. A simplification of the proposed ANN and ANFIS models.

## 4. Results and Analysis

The ANN was used to evaluate the suggested ANFIS system on the analysed topic. The ANN and ANFIS systems' findings are provided in the following subsections, respectively. The last subsection contains a comparison and explanation of these results. The MATLAB software is used to create all of the applications. To simulate and predict particles ratios, ANN and ANFIS were utilized, $\Lambda_C^+/D^0$, $\Xi_C^0/D^0$, $\Xi/\phi$ and $\Omega/\phi$[23-32], individually using experimental data. Data collected from the experiments are divided into three sets, namely, 70% training set, 15% testing set and 15% validation set. The training set is used to train the model. The validation dataset is used to confirm the accuracy of the proposed model. It ensures that the relationship between inputs and outputs, based on the training and testing sets is real.

In implementing the ANN model, with a feed-forward model, backpropagation, tansig transfer function and the LM training algorithm was the best model for simulation and predicting particle ratios, four networks (ANN1-ANN4) were training separately for particles ratios $\Lambda_C^+/D^0, \Xi_C^0/D^0, \Xi/\phi$ and $\Omega/\phi$ respectively, having 2-neurons input, one middle layer with 5, 6, 4 and 5 neurons respectively and a single-neuron output layer. Fig. 6 shows the architecture of the (ANN1-ANN4) developed in this study, including the input layer, hidden layers, and output layer. The networks were trained many times in an automatic way and stops when the best network is obtained. For effective evaluation of the accuracy of the model using MSE and R-value to determine the degree of association between the predicted and expected values. The system of equations for MSE and R-values is established in equations (1) and (2). Fig. 7 presents a number of samples, MSE, and R values for training, validation, and, testing in ANN 1, ANN 2, ANN3, and ANN 4 respectively. Fig.8 includes the structure performance evaluation of the ANNs implemented in the designated particles ratios. Fig. 8 clearly shows that the model's best validation performance is 6.0925 x $10^{-5}$, 0.000407, 0.00073457, and 8.3282 x $10^{-5}$ at epochs 5, 7, 4, and 4 for (ANN 1 – ANN 4) respectively. The enhancement of green-colored spectra after epochs 5, 7, 4, and 4 suggest the increment of mean square error (MSE), and training is halted. The regression (R) value measured the correlation between outputs and targets. R-value close to 1 indicates a close link between output and targets, as well as the model's excellent performance (see Fig. 7).

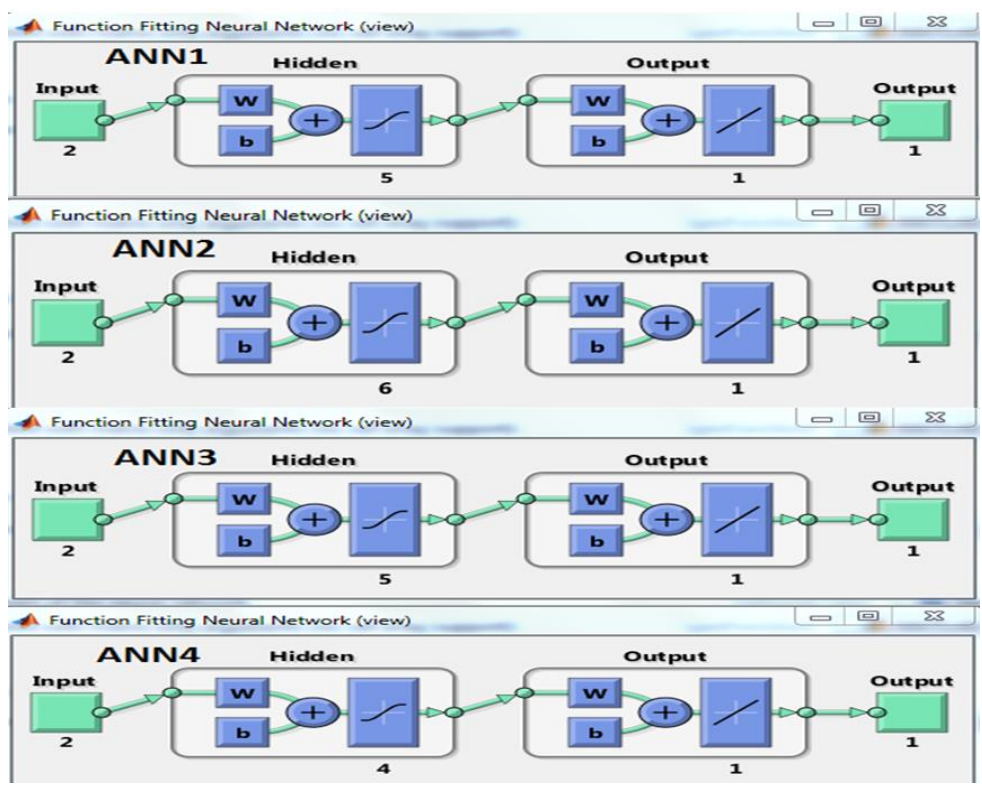

Fig. 6 Structure of the MATLAB ANN model.

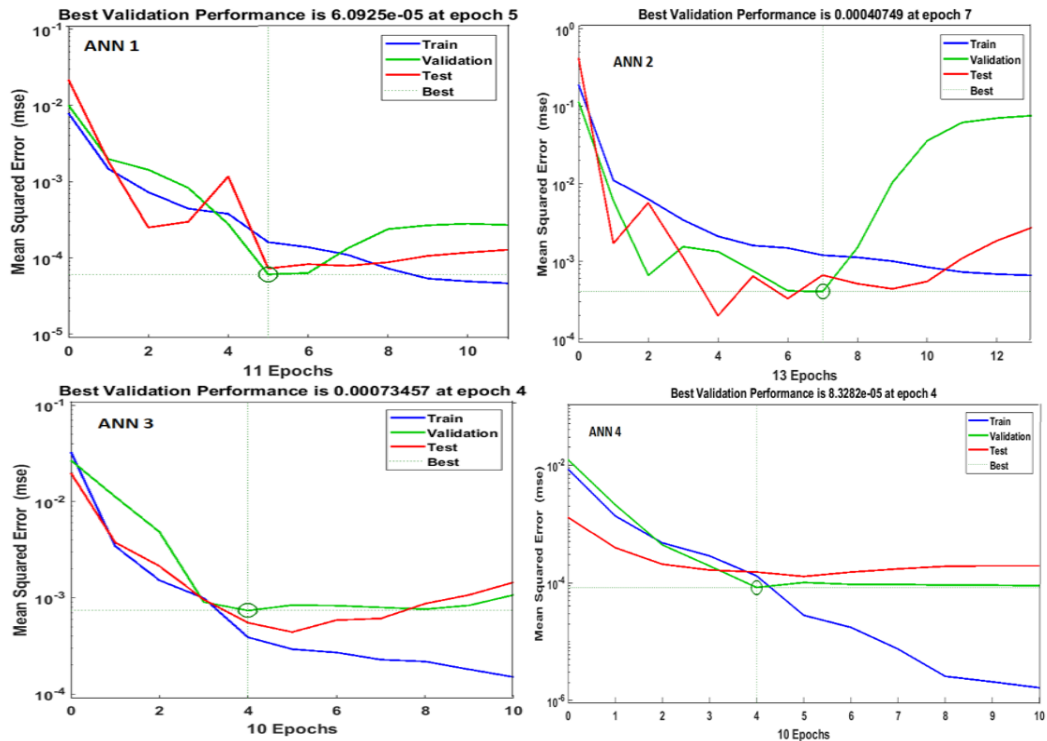

Fig. 7 Performance results of multilayer perceptron (MLP) network for different numbers.

Fig. 8 The performance of the designed ANN1 -ANN4.

The equation which describes particles ratios is given as:

particles ratios = purelin [net.LW{2,1} tansig(net.IW{1,1}( $P_T$ and $\sqrt{s}$ ) + net.b{1}) + net.b{2}]

Where
($P_T$ and $\sqrt{s}$) are the inputs
IW and LW are the linked weights as follow:
net.IW{1,1} is linked weights between the input layer and hidden layer,
net.LW{2,1} is linked weights between the hidden layer and output layer,
b is the bias and considers as follow:
net.b{1) is the bias of hidden layer,
net.b{2) is the bias of output layer.
To make the outcomes of the model reproducible, the matrices of the weights and biases for four nets of all layers of the optimized trained model are presented in appendix A.

Models with the same inputs combination as used in the ANN modeling were trained and tested to predict particles ratios using the ANFIS approach. All four models (ANFIS 1 – ANFIS 4) were trained and tested on the data of the same periods. The membership function for the models was used via trial and error technique. The hybrid learning algorithm was further used for the training of the process with an error tolerance of 0. The ANFIS 1 -ANFIS 4 were trained for 300 epochs. The ANFIS models were designed using the Trimf membership function for the ANFIS 1 and ANFIS 4 and using the Trapmf membership function for the ANFIS 2 and ANFIS 3 for inputs and linear for output. The architectures of the best trained ANFIS 1 – ANFIS 4 are shown in Fig. 9. They achieved the best MSE and higher correlation coefficient ($R$), Tab.1 captures the model specification using the adaptive neuro-fuzzy inference system (ANFIS 1- ANFIS 4).

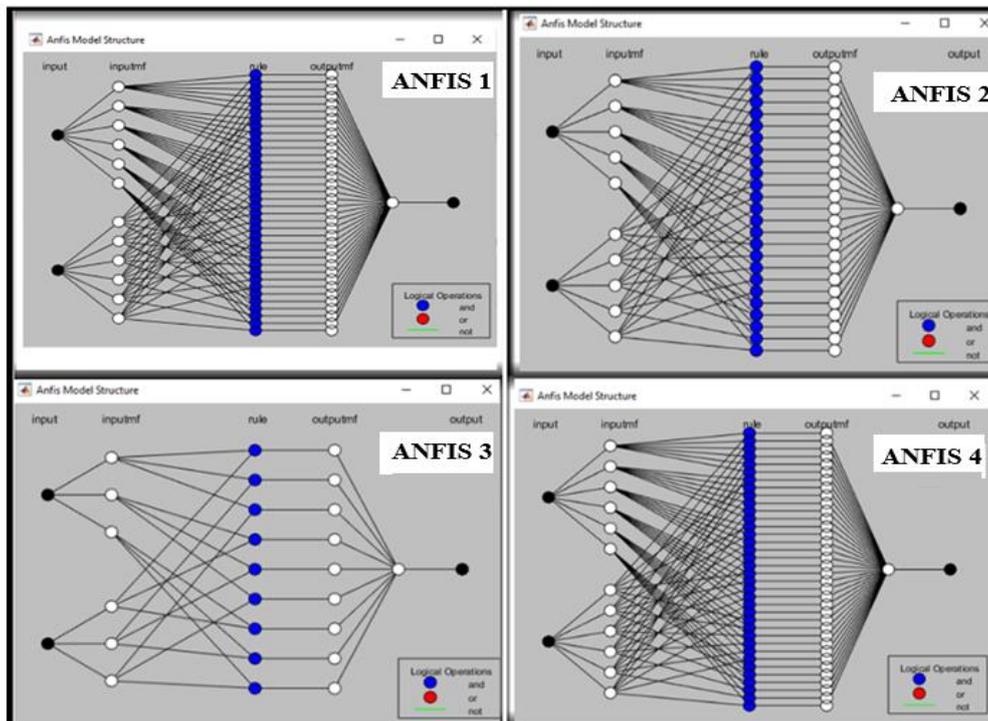

Fig. 9 The architecture of the best-trained ANFIS 1 – ANFIS 4

| ANFIS parameters type | ANFIS 1 | ANFIS 2 | ANFIS 3 | ANFIS 4 |
|---|---|---|---|---|
| Number of inputs | 2 | | | |
| Membership functions type | Trimf | Trapmf | Trapmf | Trimf |
| Number of membership functions | 6  6 | 5  5 | 3  3 | 6  6 |
| Number of nodes | 101 | 75 | 35 | 101 |
| Number of linear parameters | 108 | 75 | 27 | 108 |
| Number of nonlinear parameters | 36 | 30 | 24 | 24 |
| Total number of parameters | 144 | 105 | 51 | 132 |
| Number of fuzzy rules | 36 | 25 | 9 | 36 |
| MSE | $7.2 \times 10^{-8}$ | $3.7 \times 10^{-6}$ | 0.01 | $1.2 \times 10^{-7}$ |
| R | 0.99 | 1 | 0.978 | 0.996 |

Tab.1   Information on the best ANFIS 1- ANFIS 4 for particles ratios

After the training, they are used to predict the behavior of the p-p interaction at different values of $P_T$, and $\sqrt{s}$= 13 TeV (see Figs. 10-13). Finally, The obtained ANN and ANFIS results for particles ratios $\Lambda_C^+/D^0$, $\Xi_C^0/D^0$, $\Xi/\phi$, $\Omega/\phi$ in p-p collisions are compared to the experimental data and other theoretical results (PYTHIA (CR Mode), HERWIG7, PYTHIA, PYTHIA8 (Monash) and EPOS-LHC) at different energies ($\sqrt{s}$= 5.02 TeV(a),  $\sqrt{s}$= 7 TeV(b), and  $\sqrt{s}$= 13 TeV (c)) for particles ratios $\Lambda_C^+/D^0$ and $\Xi_C^0/D^0$ and at ($\sqrt{s}$= 5.02 TeV (a) and  $\sqrt{s}$= 13 TeV (b)) for particles ratios $\Xi/\phi$ and $\Omega/\phi$ as shown in Figs. 10 –13 respectively [33-34]. The findings produced in this work show that ANFIS and ANN models are more effective and practicable for prediction than other distributions not given in the training set and matched them precisely. The ANFIS results are slightly better than the ANN results for test and predicted stages; see Figs. 10 -13. The ANFIS system reached the highest MSE and lowest R values. It is also noticed that the ANFIS system is trained faster than the ANN system; this agreed with G. M. Behery, A. A. El-Harby and M. Y. El-Bakry [35] Khajeh and Modarress [36] and Tortum, Yayla, and Gö̈kdag̃ [37]. The major reason is that in order to get optimal performance, the ANN system executes each experiment of the same architecture numerous times. Therefore, The ANN system's training time was excessively long in comparison to the ANFIS system's training time. Additionally, if the same experiment is repeated using the ANFIS system or the ANN system, the ANFIS system will produce the same result in any iteration, however, the ANN system will produce a different result in each iteration.

Furthermore, the ANFIS system is a powerful mechanism and is trained very quickly.

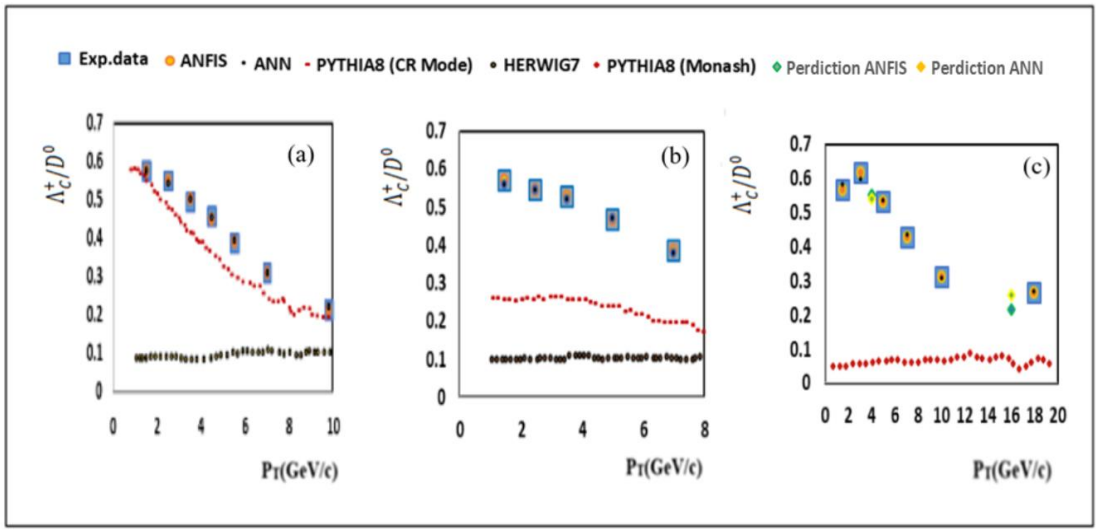

Fig. 10 $\Lambda_C^+/D^0$ ratio as a function of $P_T$ in pp collisions (a) at $\sqrt{s}$ = 5.02 TeV (b) at $\sqrt{s}$ = 7 TeV and (c) $\sqrt{s}$ = 13 TeV.

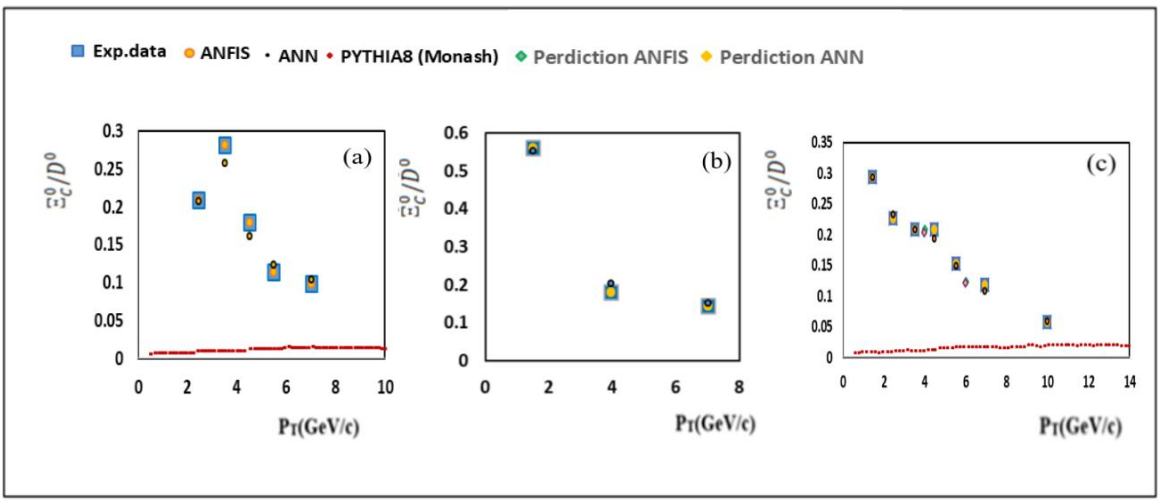

Fig. 11 $\Xi_C^0/D^0$ ratio as a function of $P_T$ in pp collisions at (a) $\sqrt{s}$ = 5.02 TeV (b) at $\sqrt{s}$ = 7 TeV and (c) $\sqrt{s}$ = 13 TeV.

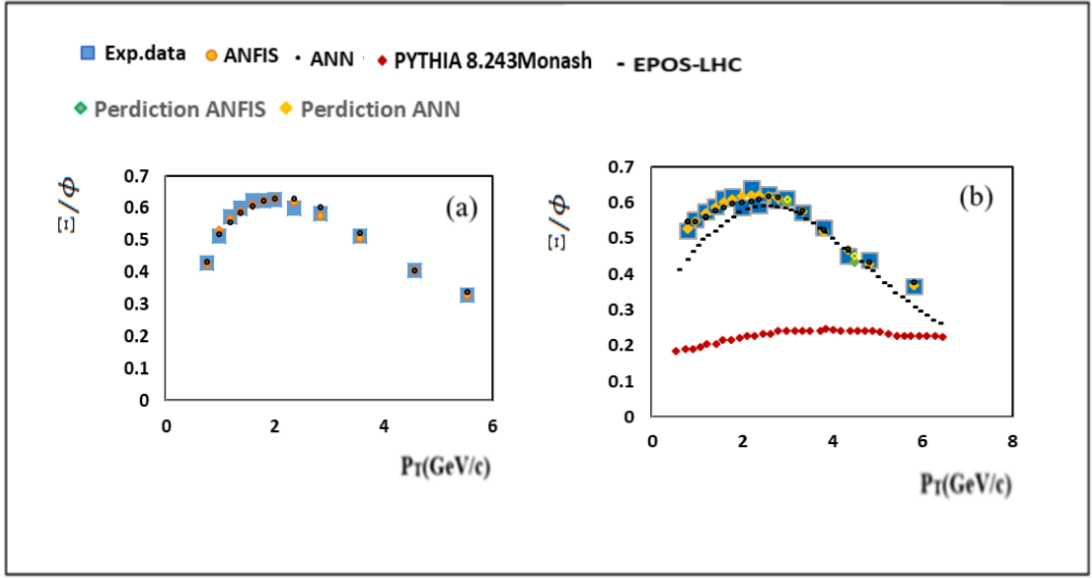

Fig.12 $\Xi/\phi$ ratio as a function of $P_T$ in pp collisions at (a) $\sqrt{s} = 7$ TeV and (b) $\sqrt{s} = 13$ TeV.

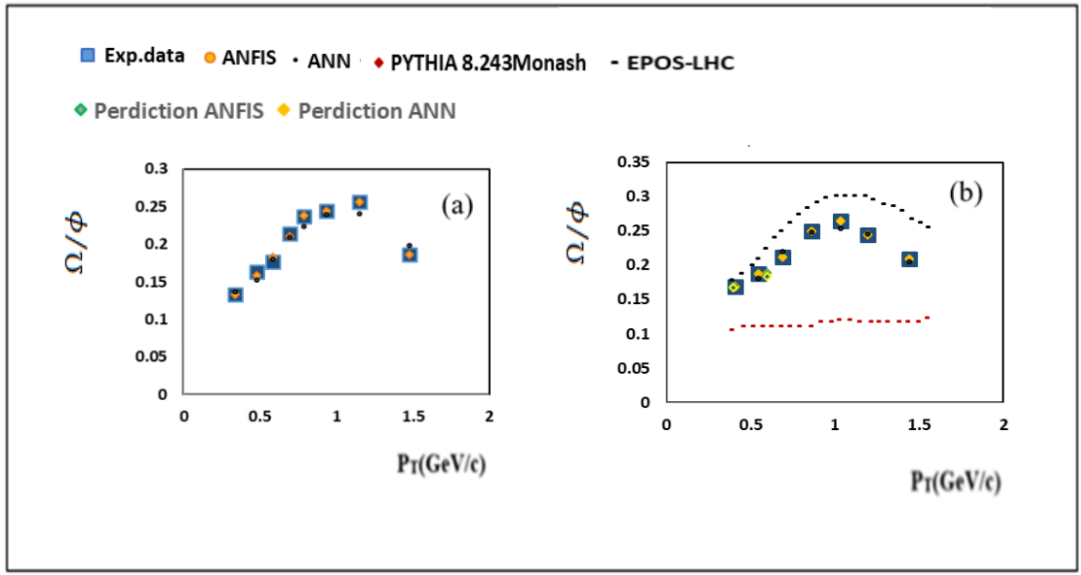

Fig.13 $\Omega/\phi$ ratio as a function of $P_T$ in pp collisions at (a) $\sqrt{s} = 7$ TeV and (b) $\sqrt{s} = 13$ TeV.

## 5. CONCLUSION

In the field of theoretical high-energy physics, the suggested ANFIS system has become well-known. By varying the kind and number of membership functions as well as the number of epochs, the system is sought to discover the optimal ANFIS that can perform the best test and prediction. As a result, various attempts are made to find the best ANFIS with the highest MSE and lowest R values. On the P-P interaction, this system and the ANN system are used and tested. The based ANFIS and ANN models calculate the particles ratios ($\Lambda_c^+/D^0$,

$\Xi_C^0/D^0$, $\Xi/\phi$, $\Omega/\phi$ ) as a function of $P_T$ and different high energies ($\sqrt{s} = 5.02$, 7 and 13 TeV ). After the training, the obtained systems predicted the values at several different values of $P_T$, and $\sqrt{s}= 13$ TeV . The ANFIS results for test and prediction sets are found to be better than the ANN results, and the ANFIS system is trained quicker than the ANN system. The ANFIS system reached the highest MSE and lowest R values. The results of simulations of training particles ratios utilising the ANFIS as evaluated using training data points demonstrated a perfect fit to the experimental data. The ANFIS' prediction capability is also good when tested with data points that were not in use during training. The results clearly show the feasibility and usefulness of such a method for obtaining collision information. The proposed ANFIS is a powerful mechanism for predicting the behavior of P-P interaction. This result may motivate us to pursue additional research. The success of the ANFIS model in describing the experimental data implies further prediction of particle ratio in the absence of experiments.

**Data Availability Statement**

All data generated and analysed during this study are included in this published article [1-6].

Appendix A:
The values of weights and bias for ANN1 was given as follows:

$$IW = \begin{pmatrix} 0.86888254415699229 & -3.3911127591123225 \\ 1.9544277908472831 & -2.5545295741405742 \\ -2.8027543811525355 & 0.23623463436845557 \\ 2.5072534043804988 & 1.510938385242411 \\ 0.70360933850406271 & 2.9843769517755852 \end{pmatrix}$$

$$LW = (\begin{matrix} -1.182701389 & 0.5243921506 & 0.8459295423 & 0.176188674856 & -0.5247446338415 \end{matrix})$$

$$b1 = \begin{pmatrix} -2.991400297374017 \\ -1.6597357893098386 \\ -1.1633242198919906 \\ 2.1040374523656187 \\ 3.0075404929459109 \end{pmatrix}$$

$$b2 = (-0.19455285578854292)$$

The values of weights and bias for ANN2 was given as follows:

$$IW = \begin{pmatrix} -0.140913421536219 & -3.6595461509518699 \\ -0.13507326616064547 & -3.384990074028384 \\ 2.8405513044077431 & 2.7470651330589417 \\ -2.1507817313348712 & 0.9989009242089254 \\ 1.4606789220241529 & -3.1635708879630897 \\ 1.6859200299020107 & -2.9090303952343604 \end{pmatrix}$$

$$LW = (\begin{matrix} 0.636995818 & 0.9116378 & 1.18447609 & 1.69763929 & 0.0962587 & -0.9709379 \end{matrix})$$

$$b1 = \begin{pmatrix} 3.1512657565580473 \\ 2.0380470471956005 \\ -1.8028665506229307 \\ -1.6970120742519403 \\ 2.5023267337254782 \\ -4.4724407220529807 \end{pmatrix}$$

$$b2 = (-0.34260982733689838)$$

The values of weights and bias for ANN3 was given as follows:

$$IW = \begin{pmatrix} 1.2787873027681869 & 2.9577409123323446 \\ -2.4617846722617474 & 1.8511933783291035 \\ -2.8496594164273352 & -1.2812135501345172 \\ -2.6548363887823099 & 0.7277023608425599 \\ 1.8559421085716081 & -2.8751702335656626 \end{pmatrix}$$

$$LW = (-0.4983198 \quad 0.5602007 \quad 0.4660032 \quad 1.36266 \quad -0.960856)$$

$$b1 = \begin{pmatrix} -3.194658411068569 \\ 1.760324818943277 \\ -0.28699863151354227 \\ -1.9873688632235869 \\ 2.9360018644215642 \end{pmatrix}$$

$$b2 = (-0.73190976568010835)$$

The values of weights and bias for ANN4 was given as follows:

$$IW = \begin{pmatrix} 2.7204629321420475 & 0.45811841483546112 \\ -0.76723465375418232 & -2.9624070275661039 \\ 3.3511401268988692 & -0.15793492142740212 \\ -1.9556106455698008 & 1.0045359489339909 \end{pmatrix}$$

$$LW = (\ -0.62560886327842202 \quad -0.20529710541534871 \quad 0.76151656733351858 \quad 0.46565132835516815)$$

$$b1 = \begin{pmatrix} -2.7203003401662609 \\ 1.9670257450444841 \\ 1.6957514404535012 \\ -3.3527494406772691 \end{pmatrix}$$

$$b2 = (-0.21306246371306878)$$